\documentclass[a4paper]{jpconf}
\usepackage{graphicx}
\usepackage{amsmath} 

\newcommand{\kms}{\rm km\ s^{-1}}
\newcommand{\ergs}{\rm erg\ s^{-1}}
\newcommand{\flux}{\rm erg\ s^{-1}\ cm^{-2}}
\newcommand{\mic}{\mbox{$\mu$m}}

\newcommand{\kev}{\rm keV}

\let\AAold\AA
\renewcommand{\AA}{\text{\AAold}}

\newcommand{\lbol}{L_{\rm bol}}
\newcommand{\lledd}{L / L_{\rm{Edd}}}
\newcommand{\mbh}{M_{\rm BH}}
\newcommand{\an}{\alpha_{\rm opt}}
\newcommand{\nln}{\nu L_{\nu}}
\newcommand{\nlnl}{\nu L_{\nu}}
\newcommand{\Ks}{K$_{\rm s}$}

\newcommand{\msun}{{\rm M_{\odot}}}

\newcommand{\mgii}{\text{Mg~{\sc ii}}}
\newcommand{\civ}{\text{C~{\sc iv}}}

\newcommand{\Ha}{\text{H$\alpha$}}

\newcommand{\Hb}{\text{H$\beta$}}

\newcommand{\lbha}{L_{\rm bH\alpha}}

\newcommand{\dv}{\Delta {\rm v}}
\newcommand{\lagn}{L_{\rm AGN}}
\newcommand{\lhost}{L_{\rm host}}
\newcommand{\lx}{L_{\rm X}}

\renewcommand{\th}{$^{\rm th}$}

\newcommand{\psmpsz}{232\,837}
\newcommand{\smpsz}{3\,579}

\newcommand{\aap}{{\it A\&A}}
\newcommand{\apj}{{\it ApJ}}

\newcommand{\apjs}{{\it ApJS}}
\newcommand{\aj}{{\it AJ}}
\newcommand{\mnras}{{\it MNRAS}}

\begin{document}

\title[Type 1 low $z$ AGN]{Type 1 AGN at low $z$} 
\author[Jonathan Stern and Ari Laor]
{J Stern and A Laor}

\address{Department of Physics, Technion -- Israel Institute of Technology, Haifa, Israel}

\ead{stern@physics.technion.ac.il, laor@physics.technion.ac.il}

\begin{abstract}
We present the emission properties of a sample of \smpsz\ type 1 AGN, selected based on the detection of broad \Ha\ emission. The sample covers the range of black hole mass $10^6<\mbh/\msun<10^{9.5}$ and luminosity in Eddington units $10^{-3} < \lledd < 1$. 
Our main results are: 
1. The distribution of the \Ha\ FWHM values is independent of luminosity.
2. The observed mean optical-UV SED is well matched by a fixed shape SED of luminous quasars, which scales linearly with broad \Ha\ luminosity, and a host galaxy contribution. 
3. The host galaxy $r$-band (fibre) luminosity function follows well the luminosity function of inactive non-emission line galaxies (NEG), consistent with a fixed fraction of $\sim 3$\% of NEG hosting an AGN, regardless of the host luminosity. 
4. The optical-UV SED of the more luminous AGN shows a small dispersion, consistent with dust reddening of a blue SED, as expected for thermal thin accretion disc emission. 
5. There is a rather tight relation of $\nln(2\ \kev)$ and broad \Ha\ luminosity, which provides a useful probe for unobscured (true) type 2 AGN.
\end{abstract}

\section{Introduction}

We examine the emission properties of a new sample of low $z$ broad line AGN, aimed to be an extension of the SDSS quasar catalog (QCV, Schneider et al. 2010) to low luminosities. This contribution summarizes the main results obtained on the distribution of the broad line FWHM (\S3), the host galaxies (\S4), and the AGN spectral energy distribution (\S\S5--7). Full details of the methods used, and a discussion of the implication of the results, are given in the main article (Stern \& Laor, accepted).

\section{Sample Creation}

The type 1 (T1) sample is selected from the SDSS 7\th\ data release (DR7, Abazajian et al. 2009), based on the detection of a broad \Ha\ emission line. The parent sample includes \psmpsz\ spectra with S/N $>10$ and $0.005<z<0.31$. The broad \Ha\ emission is identified by modeling the nearby narrow emission lines, stellar absorption features and featureless continuum. We find broad \Ha\ emission in \smpsz\ objects (1.5\%), which constitute the T1 sample. 

The optical spectra in the T1 sample are supplemented by nearly complete near IR photometry from 2MASS (Skrutskie et al. 2006, detection rate 97\%) and UV photometry from GALEX  (Martin et al. 2005, 93\%). We also add X-ray photometry from the ROSAT (Voges et al. 1999) survey, available for 43\% of the T1 objects. The measured broad \Ha\ luminosities ($\lbha$) and FWHMs ($\dv$) and the photometric luminosities are available electronically.

\section{How broad do the broad lines get?}
Various theoretical models of the broad line region (BLR) and the AGN ionizing continuum predict an upper limit on $\dv$, beyond which the BLR does not exist (Nicastro 2000, Elitzur and Shlosman 2006, Laor \& Davis 2011). The models differ in the dependence of the maximal $\dv$ on luminosity. In Figure 1, we plot the $\lbha$ vs. $\dv$ distributions of the T1 objects and the 8\,185 QCIV objects where the broad \Hb\ is available (measurements from Shen et al. 2008). 

\begin{figure}
\begin{center}
\includegraphics{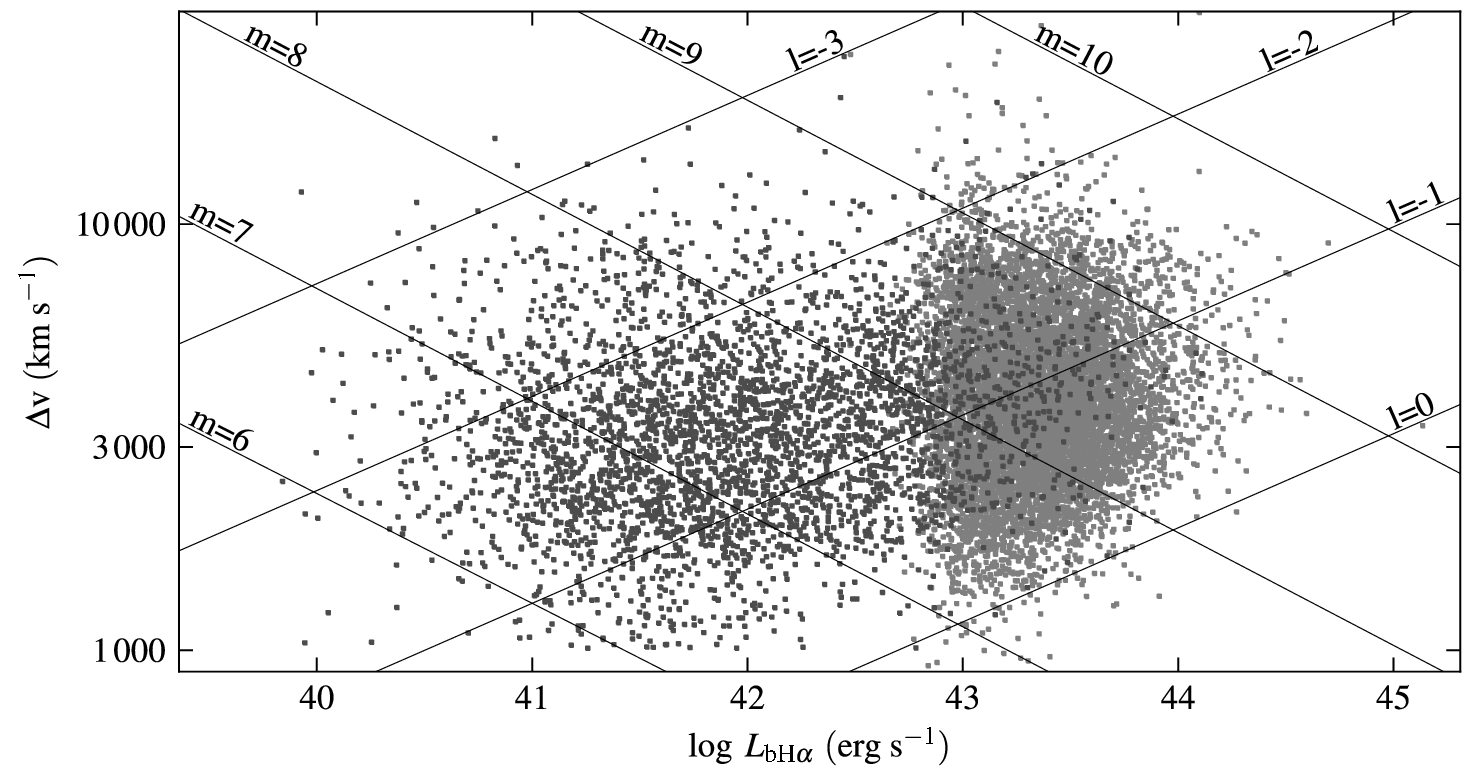}
\end{center}
\caption{\label{label}The distributions of the T1 (dark gray) and QCIV (light gray) samples in the broad \Ha\ luminosity vs. FWHM plain. Crossing lines mark decades of the implied $\mbh$ and $\lledd$.}
\end{figure}

The T1 sample extends down by two orders of magnitude in $\lbha$ compared to QCIV. The Eddington ratio $\lledd\ (\equiv 10^l)$ and black hole mass $\mbh\ (\equiv 10^m\msun)$ are derived from $\lbha$ and $\dv$, in a manner similar to Greene \& Ho (2005). It can be seen that the Eddington limit ($l=0$) sets a minimum $\dv$ with increasing $\lbha$ for $\lbha > 10^{42.5}\ \ergs$. This, together with the rarity of AGN with $m>9.5$, leads to a decrease in the range of observed $\dv$ values with increasing $\lbha$. A similar convergence of the range of $\dv$ for the \mgii\ and \civ\ lines at the highest continuum luminosities was noted by Fine et al. (2008, 2010).

There is a steep decline in the number of objects with $\dv>10\,000\ \kms$, at all luminosities. This decline is seen clearly in Figure 2, which plots the $\dv$ distribution of the combined T1 + QCIV sample at different $\lbol$. The distributions are remarkably similar, showing a roughly linear decline of $\log ({\rm d} N/{\rm d}\log \dv)$ vs. $\dv$, or equivalently ${\rm d} N/{\rm d}\log \dv \propto e^{-\dv/\dv_0}$, with $\dv_0 \approx 2700\ \kms$. The origin of this similarity is not clear, as the $\dv$ distributions should be set by the distribution of the $m$ and $l$ values. Either these distributions somehow lead to a $\dv$ distribution which is independent of $\lbha$, or it may imply that $\dv$, rather than $m$ and $l$, sets the observed $\dv$ distribution through some unknown mechanism. Note the effect of the Eddington limit at $\dv < 3\,000\ \kms$, which increases the minimal $\dv$ and the peak position with increasing $\lbol$.

\begin{figure}
\begin{minipage}{18pc}
\includegraphics[width=17pc]{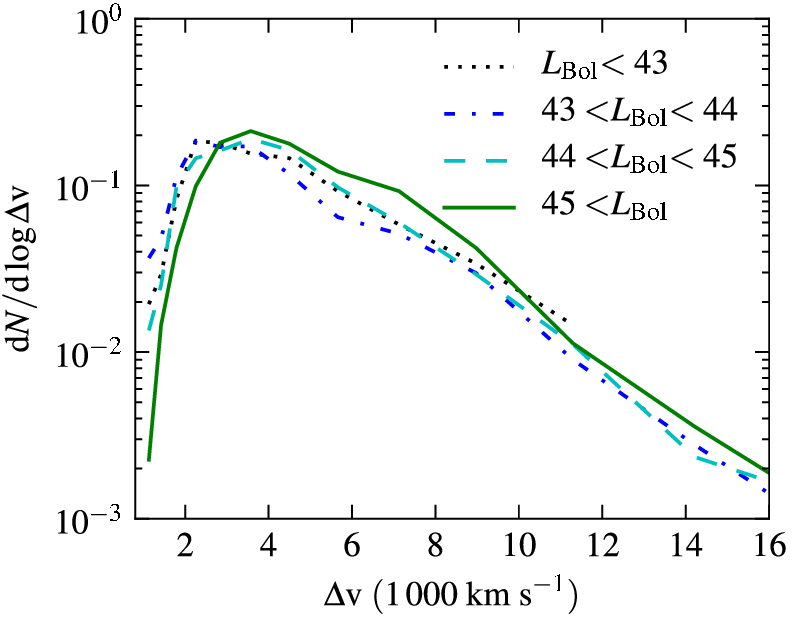}
\caption{The FWHM distribution of the T1+QCIV sample, at different $\lbol$ bins.}
\caption{The spectral type distribution of galaxies in the parent sample, as a function of luminosity ($r$ band). Non-T1 classifications are from Brinchmann et al. (2004). The `T1 host' line is corrected for the AGN contribution. The lower panel shows the number of objects per 0.25 decade in $\nln$.}
\end{minipage}\hspace{2pc}
\begin{minipage}{18pc}
\includegraphics{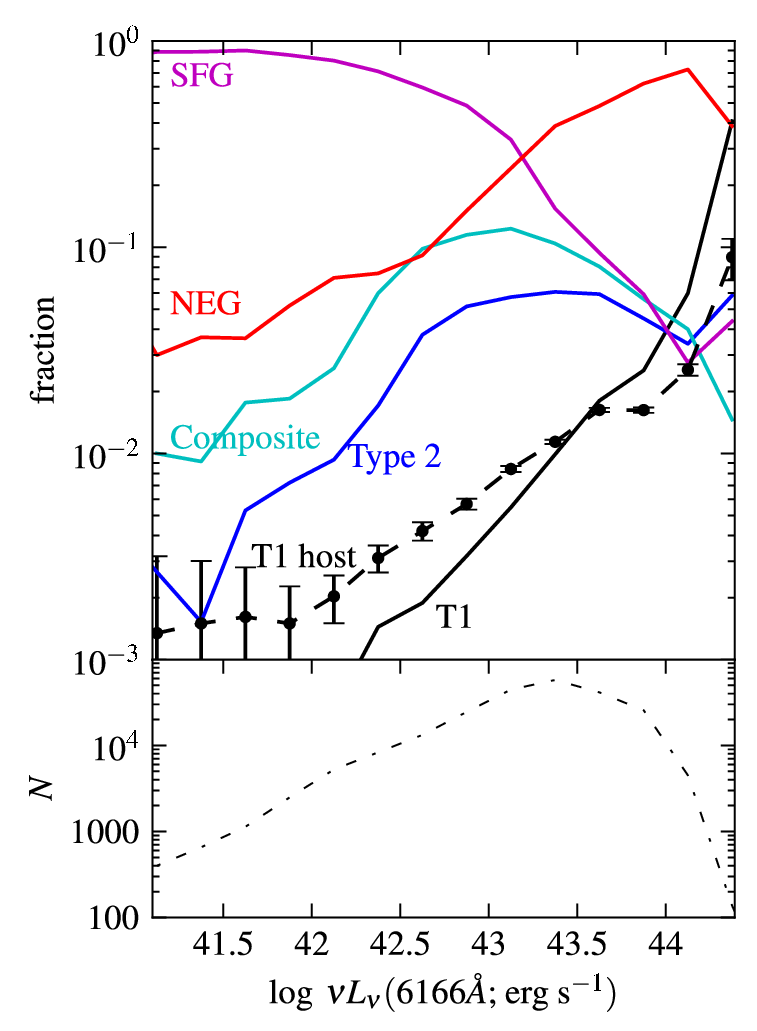}
\end{minipage} 
\end{figure}

\section{How does the AGN fraction change with galaxy luminosity?}

Kauffmann et al. (2003), and various follow up studies, found that AGN tend to reside in massive hosts. We therefore examine the T1 fraction as a function of host luminosity.

Figure 3 presents the classification of $z<0.2$ galaxies in the parent sample (\S2) as a function of the spectroscopically measured $\nln$(6166\AA). The type 2, Star Forming Galaxy (SFG), Non-Emission line Galaxy (NEG) and composite classifications (based on the equivalent width and BPT ratios of the narrow lines) are taken from Brinchmann et al (2004). 
At low luminosities $>$90\% of the SDSS galaxies are SFG, while at high luminosities the NEG dominate. The `T1 host' line is derived from the $\nln$(6166\AA) of the T1 objects, corrected for the AGN contribution (see \S5 below). Remarkably, the T1 host fraction follows well the NEG fraction, or equivalently, their galaxy luminosity distribution is similar. This may suggest that T1s reside in NEG, as also implied by their similar concentration indices and bulge/total light ratios (Kauffmann et al. 2003). The fraction of NEG hosting broad line AGN at the level detectable in this study is $\sim 3\%$, independent of galaxy luminosity.

\section{How does the AGN spectral energy distribution (SED) depend on luminosity?}

Figure 4 compares $\lbha$ with the observed 2.2 \mic\ -- 2 \kev\ continuum luminosity, in the T1 sample.
In the luminous bins in the FUV, NUV and 3940\AA\ panels, the slopes are consistent with a linear relation, in all other bands and luminosities the slopes are flatter. Since the FUV occurs close to the peak position of the AGN SED (e.g. Zheng et al. 1997), the linear relation between $\nlnl$(FUV) and $\lbha$ suggests that $\lbha$ provides a good estimator of $\lbol$. Also, the linear relation suggests that the mean covering factor of the BLR is independent of luminosity.

\begin{figure}
\includegraphics[width=38pc]{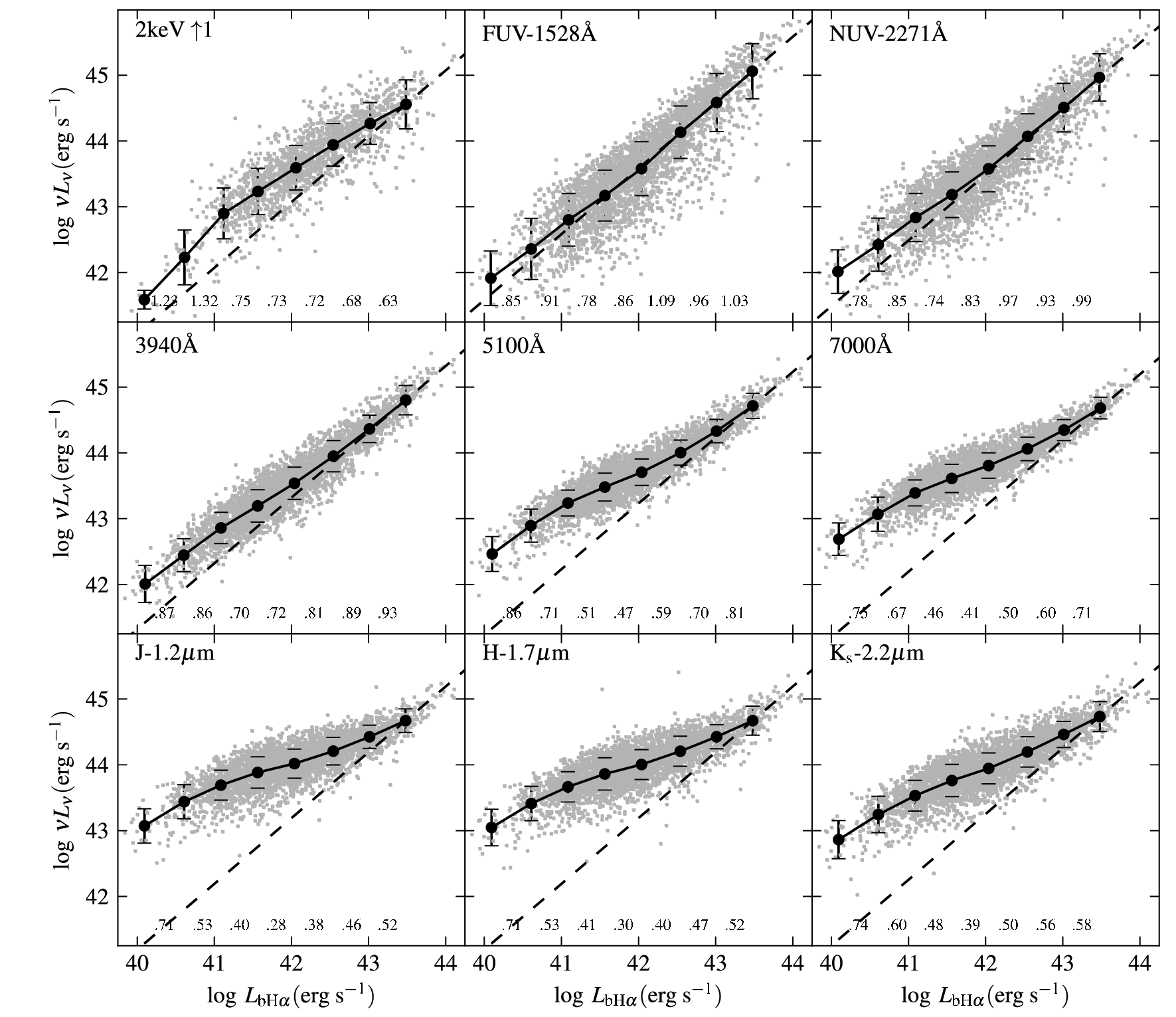}
\caption{
The $\lbha$ vs. observed (AGN+host) $\nlnl$ from the X-ray to the NIR, in the T1 sample. The X-ray $\nlnl$ are shifted upward by 1 decade for presentation purpose.
The large black dots and associated error bars mark the mean and dispersion of $\nln$ in half-decade wide $\lbha$ bins.
The solid lines connect the means of adjacent bins, with the local slopes written underneath. Dashed lines depict a linear relation, normalized to the highest $\lbha$ bin.
} 
\end{figure}

In all bands excluding X-ray, there is a transition from a steeper slope to a flatter slope with decreasing $\lbha$. The transition $\lbha$ increases with wavelength, up to a maximum of $\simeq 10^{43}\ \ergs$ at 1.2\mic. The transition occurs due to the host contribution, which peaks at 1.2\mic. 

The transition between AGN and host dominance of the SED can be seen in Figure 5. The left panel plots the mean SEDs in the $\lbha$ bins, i.e. each line connects the mean values of different continuum wavelengths for the same $\lbha$ (Fig. 4). Note the transition from an AGN-dominated UV peaked SED to a galaxy-dominated NIR peaked SED with decreasing $\lbha$. As the net AGN SED, we take the Richards et al. (2006, hereafter R06) SED derived from luminous quasars in the SDSS. We scale the R06 SED to the UV in the highest $\lbha$ bin, and rescale by the relative $\lbha$ of the lower seven bins. The scaled R06 SEDs are the putative net AGN SEDs, assuming the mean SED remains fixed with luminosity and scales linearly with $\lbha$. In the middle panel of Fig. 5 we plot the residuals derived by subtracting the AGN from the observed SED, and in the right panel we plot the mean SEDs of inactive SDSS galaxies, with $z$-distributions matched to the $z$-distributions of each $\lbha$ bin.

The residuals all appear to have a typical galaxy SED. This justifies (qualitatively) the simple scaling prescription we used for the net AGN SED. Note the significant excess luminosity in the \Ks\ and H bands in the most luminous residuals, and the non-linear $\nln$(X-ray)-$\lbha$ relation (Fig. 4), which imply that the linear scaling of the net AGN SED is applicable only to the optical-UV region.

\begin{figure}
\begin{center}
\includegraphics{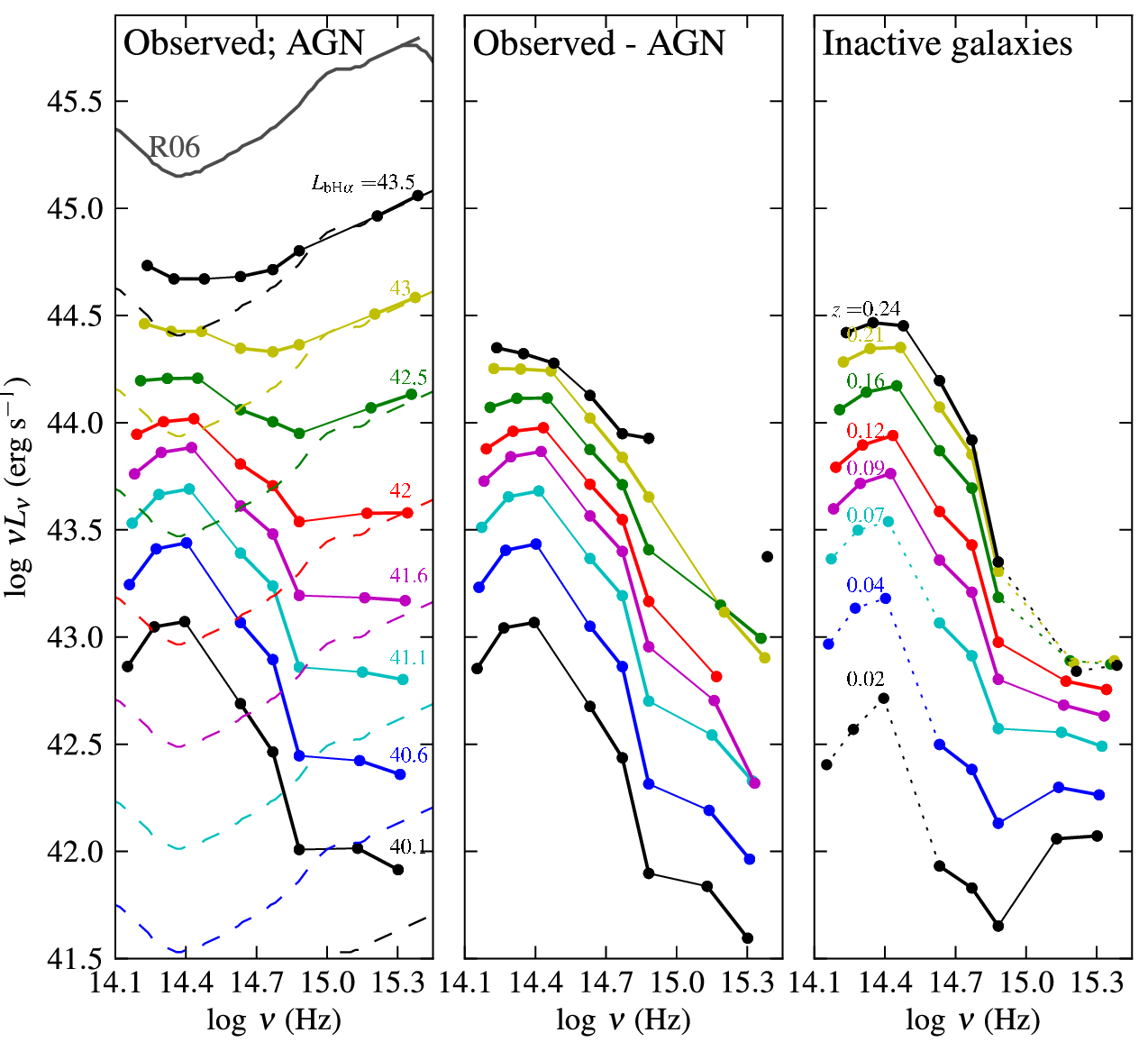}
\caption{The luminosity dependence of the mean T1 objects SEDs at 2.2\mic\ -- 1528\AA. {\bf Left} Solid lines mark the observed SEDs for the eight $\lbha$ bins (mean $\lbha$ noted). Dashed lines mark the net AGN SED, derived by scaling the R06 quasar SED by the mean $\lbha$ of the bin with the same color. {\bf Middle} Residuals of the subtraction of the the scaled R06 SEDs from the observed AGN SEDs. {\bf Right} Mean SEDs of $z$-matched inactive galaxies (mean $z$ noted). Dotted lines mark bands with $<70\%$ detections.
} 
\end{center}
\end{figure}

In order to quantitatively evaluate the AGN SED scaling law, we define the scaling index $\beta$, where $\lagn(\lambda) \propto \lbha^{\beta}$. A $\beta$ independent of wavelength $\lambda$ indicates the mean AGN SED shape remains fixed with luminosity. From the qualitative analysis shown in Fig. 5, we expect $\beta \sim 1$ for $\lambda$ in the optical or UV. 
In Figure 6, we plot the implied mean host luminosity $\lhost \equiv \nln - \lagn$ in the SDSS $u$ and $z$ bands, for different $\lbha$ and $\beta$. As the expected mean $\lhost$ of a certain $\lbha$ bin, we use the mean luminosity of type 2 AGN with the same $z$ distribution as the $\lbha$ bin. The type 2s are taken from the sample of Brinchmann et al. (2004). 

The top panel of Fig. 6 shows that in the $u$ band, $\beta < 0.9$ is ruled out, as the implied residual is negative for $\lbha > 10^{41}\ \ergs$. 
It is remarkable that the simplest scaling, $\beta=1$ in both bands, leads to the closest match of the mean type 1 host to the mean type 2 host. Together with the linear $\nln$(UV)-$\lbha$ relationship at high luminosities (Fig. 4), this implies that the mean optical-UV net AGN SED has a fixed shape, and scales linearly with $\lbha$. 

\begin{figure}
\begin{minipage}{18pc}
\includegraphics{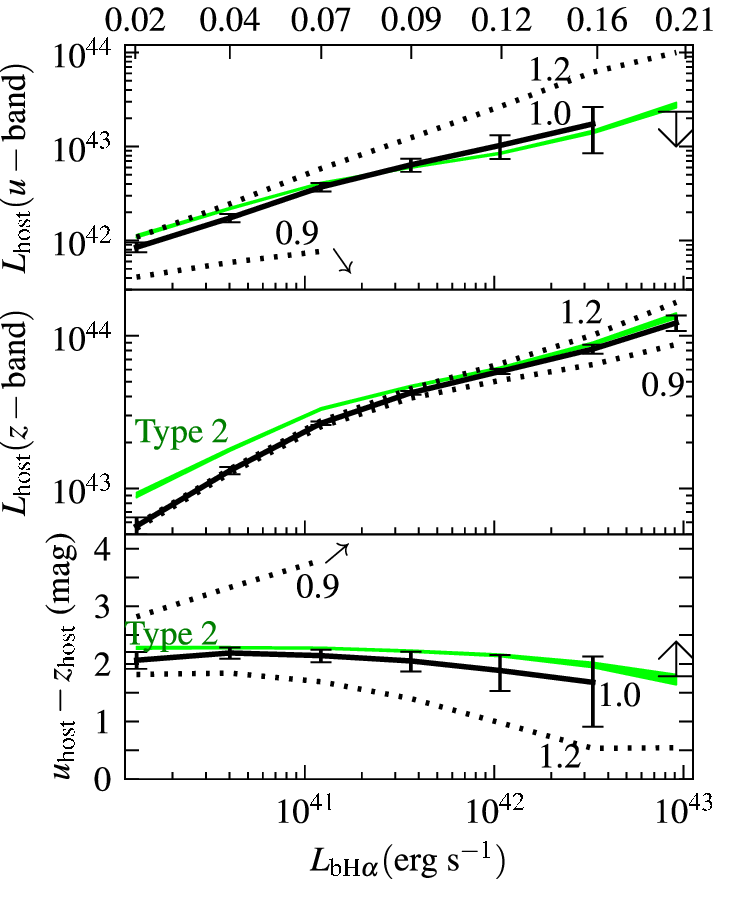}
\caption{\label{label} A comparison of the implied mean T1 host luminosity for different $\lbha$ bins, for different AGN SED scaling laws (black solid/dotted lines, $\beta$ noted), with $z$-matched type 2 AGN. The comparison is made in the $u$ band ($3551\AA$), $z$ band ($8932\AA$), and $u-z$ color. The mean $z$ of each $\lbha$ bin is noted on top. Arrows mark a negative implied host.
}
\end{minipage}\hspace{2pc}
\begin{minipage}{18pc}
\includegraphics{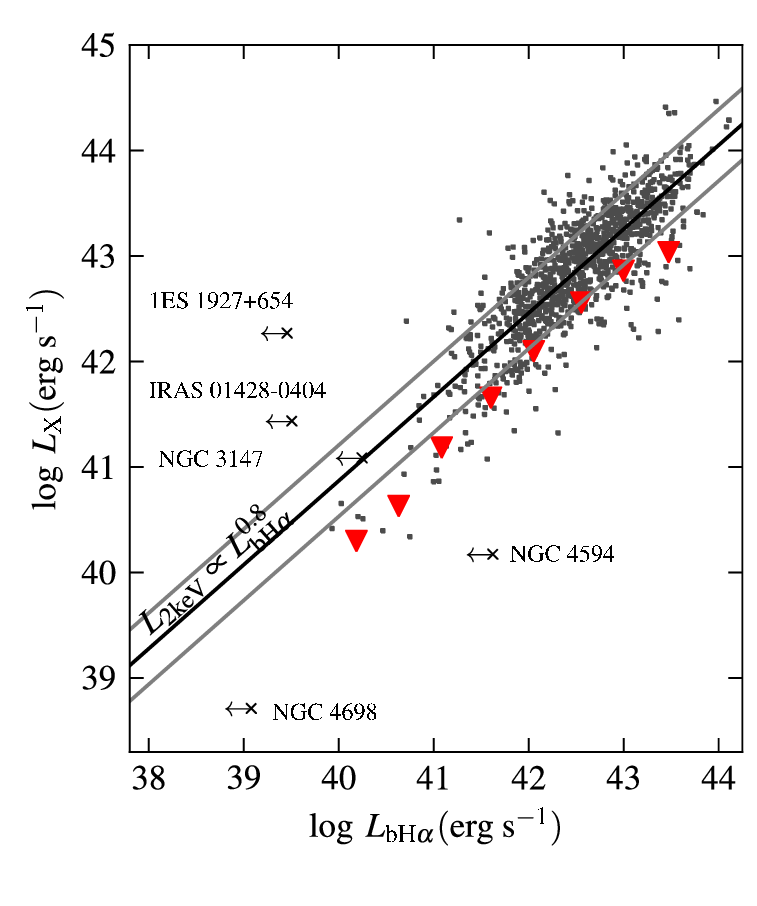}
\caption{The $\lx$ vs. $\lbha$ relation, as a probe of true type 2 AGN. Dots mark T1 objects with $F_{\rm b\Ha} >10^{-13.5}\ \flux$. Red triangles mark mean upper limits in 0.5-dec $\lbha$ bins. The best power law fit and dispersion are shown as black and gray lines (slope noted). Arrows mark the upper limits on $\lbha$ of five true type 2 candidates.}
\end{minipage}
\end{figure}

\section{Is the lack of BLR emission in true type 2 candidates significant?}
True type 2 candidates are AGN in which the BLR is not observed, despite apparently unobscured X-ray emission. In Figure 7 we compare the upper limits on $\lbha$ in five true type 2 candidates from Shi et al. (2010) and Tran et al. (2011), with the T1 sample $L_{\rm X} (\equiv \nln(2\kev))$ vs. $\lbha$ relation. Upper limits on $\lbha$ are derived from the expected $\dv$ (based on the published $\mbh$ and $\lledd$, \S3) and the flux density near \Ha. The absence of a broad \Ha\ in NGC~4594, NGC~4698, and NGC~3147 is not significant. In the other two objects the expected $\lbha$ is well above the upper limits, and these two objects appear to be true type 2 AGN.

\section{What drives the dispersion in the optical-UV SED of AGN?}

In \S5 we have shown that the mean SED of broad line AGN is well reproduced by the sum of the mean SED of luminous quasars, scaled down by $\lbha$, and a host contribution. Here, we examine the dispersion in the shapes of individual SEDs.

In Figure 8, we show the distributions of $\nln$(FUV)$/\lbha$ and optical slope $\an$ as a function of $\lbha$. 
The $\nln$(FUV)$/\lbha$ distributions for $\log \lbha>40.5$ have a very similar shape. All distributions have a narrow peak, with an extended tail towards low $\nln$(FUV)/$\lbha$ values. The narrow peak suggests there is a small dispersion in the covering factor of the BLR. In the $\log \lbha=40.5$ bin, host contribution to the UV broadens the distribution and increases the mean value. 

\begin{figure}
\begin{minipage}{18pc}
\includegraphics{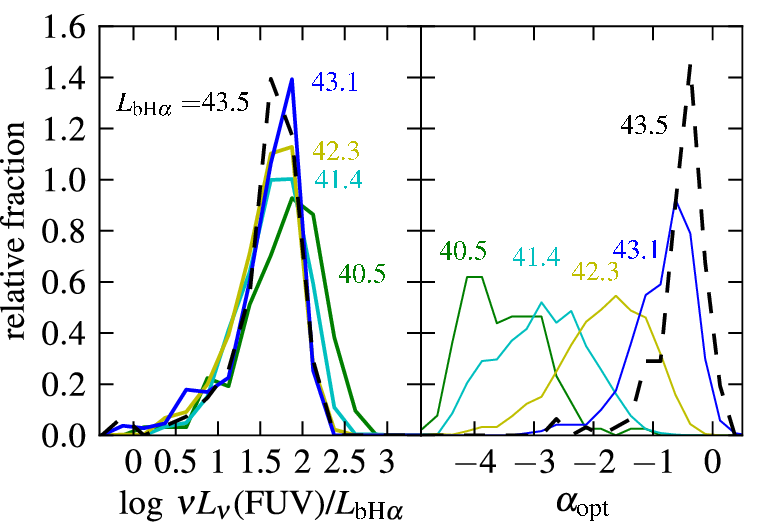}
\caption{The T1 objects $\nln$(FUV)$/\lbha$ and $\an$ distributions as a function of $\lbha$. Solid (dashed) lines represent bins of width 1 (0.5) dec in $\lbha$. The mean $\lbha$ are noted.}
\end{minipage}\hspace{2pc}
\begin{minipage}{18pc}
\includegraphics[width=17pc]{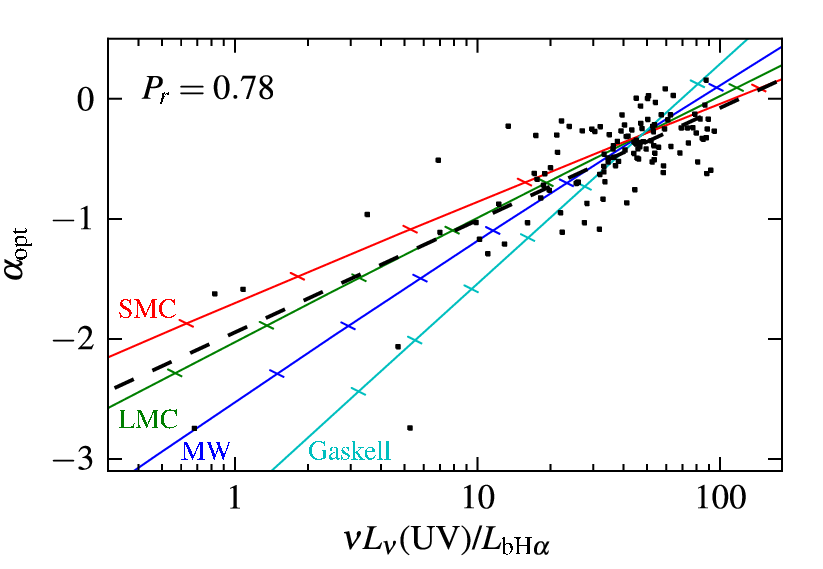}
\caption{The $\an$ vs. $\nln$(UV)/$\lbha$ of the T1 objects in the $\log\ \lbha=43.5$ bin. Also shown are the best fit (dashed line) and expected relations for different extinction laws (solid lines, marks every 0.1 in $E_{\rm B-V}$). }
\end{minipage} 
\end{figure}

The $\an$ distributions are redder at lower luminosities, due to the host emission. The $\log\ \lbha = 43.5$ bin is least affected by the host contribution, and has a narrow peak and extended red tail, similar to the $\nln$(FUV)$/\lbha$ distribution, and to the continuum slope distribution found by Richards et al. (2003) on luminous quasars. There is therefore a small dispersion in $\nln$(FUV)/$\lbha$ and $\an$ in the net AGN emission, based on the objects least affected by the host contribution. The similar shape may indicate a common origin of the dispersion, such as dust reddening. 

The dust reddening scenario is explored in Figures 9 and 10. In Fig. 9 we plot $\an$ vs. $\nln$(FUV)/$\lbha$ for the $\log \lbha=43.5$ bin. The two independent ratios are highly correlated, as expected if dust extinction contributes to the dispersion of the AGN SED. An SMC/LMC extinction law (using the Pei 1992 formulation) is preferred over a Milky Way or Gaskell \& Benker (2007) extinction law, as found by Hopkins et al. (2004) on luminous quasars. 
In Fig. 10, we plot the 2.2 \mic\ -- 2 \kev\ mean SED for different $\an$ bins at the highest $\lbha$ bins. In the $\log \lbha=43.5$ panel, the difference between SEDs increases from the NIR to the UV and disappears in the X-ray, also consistent with dust extinction. A similar behavior is observed in the $\log\ \lbha=43$ panel, together with a weak anti-correlation of NIR luminosity and $\an$. The latter trend originates from the range of host luminosities, where a relatively luminous host increases $\nln$(1.2\mic) and decreases $\an$.

The reddening origin for the dispersion in the net AGN SED implied by Figs. 9 and 10, already at $\an>-1$, indicates that the intrinsic AGN SED dispersion is even smaller than seen in Fig. 8. A rather uniform and blue intrinsic optical-UV SED is expected if AGN are powered by a thin accretion disc, which emits locally close to a blackbody.

\begin{figure}
\includegraphics{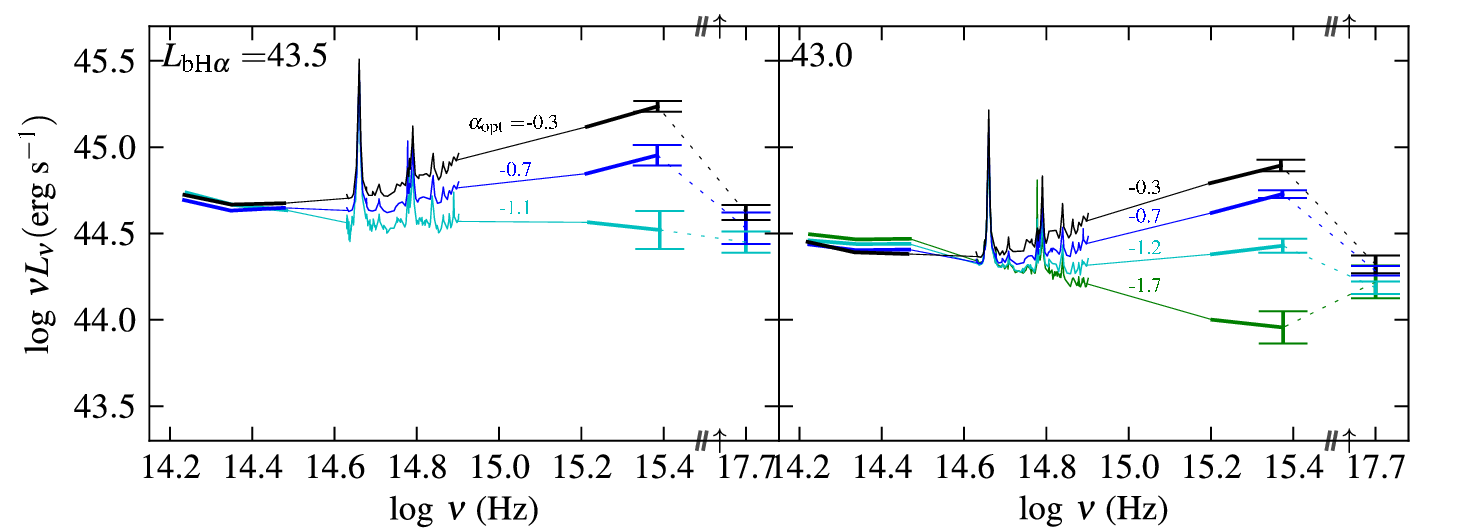}
\caption{The mean IR to X-ray SEDs and optical spectra as a function of $\an$, for T1 objects with $\lbha = 10^{43.5}$ and $10^{43}\ \ergs$. The value of $\lx$ is shifted upward (+1 dec) for presentation. The error bars denote the errors in the mean $\lx$ and $\nln$(FUV). 
}
\end{figure}

\section{Conclusions}
We present and analyze a new sample (T1) of \smpsz\ broad \Ha\ selected AGN from the SDSS DR7, with $\log \lbha=40-44$, which spans $m=6-9$ and $l=-3-0$. We add UV (GALEX), IR (2MASS), and X-ray (ROSAT) luminosities to form the mean SED. The main results are:

\begin{enumerate}
\item The \Ha\ FWHM velocity distribution ${\rm d} N/{\rm d}\log \dv$ is independent of luminosity and falls exponentially with $\dv$. 
The origin of this distribution remains to be understood.
\item The observed mean 9000\AA--1500\AA\ SED, as a function of $\lbha$, is consistent with a sum of the mean SED of luminous quasars, which scales linearly with $\lbha$, and a host galaxy contribution.
\item The host galaxy $r$-band luminosity function of T1 objects follows the NEG luminosity function, with a relative normalization of $\sim 3$\%, suggesting that the host of broad line AGN are NEG, and the AGN probability of occurrence is independent of the host mass. 
\item The dispersion in the optical-UV SED in luminous AGN ($\log \lbha \ge 43$), is consistent with reddening. 
This indicates the intrinsic SED of AGN is blue, with a small dispersion, as predicted from thermal thin accretion disc models.
\item The $\lbha$ versus $L_{\rm{X}}$ correlation provides a useful probe for unobscured narrow line AGN. It can be used to test if the absence of a broad \Ha, in X-ray detected AGN, is significant.
\end{enumerate}

\section*{References}

 \begin{thereferences}
\item Abazajian K~N \etal 2009 \apjs\ {\bf 182} 543 
\item Brinchmann J, Charlot S, White S~D~M, Tremonti C, Kauffmann G, Heckman T and Brinkmann J 2004 \mnras\ {\bf 351} 1151
\item Elitzur M and Shlosman I 2006 \apj\ {\bf 648} L101 
\item Fine S \etal 2008 \mnras\ {\bf 390} 1413
\item Fine S \etal 2010 \mnras\ {\bf 409} 591 
\item Gaskell C~M and Benker A~J 2007 {\it ArXiv e-prints} 0711.1013
\item Greene J~E and Ho L~C 2005 \apj\ {\bf 630} 122 
\item Hopkins P~F \etal 2004 \aj\ {\bf 128} 1112 
\item Kauffmann G \etal 2003 \mnras\ {\bf 346} 1055
\item Laor A and Davis S~W 2011 \mnras\ {\bf 417} 681 
\item Martin D~C \etal 2005 \apj\ {\bf 619} L1 
\item Nicastro F 2000 \apj\ {\bf 530} L65 
\item Pei Y~C 1992 \apj\ {\bf 395} 130 
\item Richards G~T \etal 2003 \aj\ {\bf 126} 1131 
\item Richards G~T \etal 2006 \apjs\ {\bf 166} 470 (R06)
\item Schneider D~P \etal 2010 \aj\ {\bf 139} 2360 
\item Shen Y, Greene J~E, Strauss M~A, Richards G~T and Schneider D~P 2008 \apj\ {\bf 680} 169 
\item Shi Y, Rieke G~H, Smith P, Rigby J, Hines D, Donley J, Schmidt G and Diamond-Stanic A~M 2010 \apj\ {\bf 714} 115 
\item Skrutskie M~F \etal 2006 \aj\ {\bf 131} 1163 
\item Stern J and Laor A, accepted for publication in {\it MNRAS}, {\it ArXiv e-prints} 1203.3158
\item Tran H~D, Lyke J~E and Mader J~A 2011 \apj\ {\bf 726} L21 
\item Voges W \etal 1999 \aap\ {\bf 349} 389 
\item Zheng W, Kriss G~A, Telfer R~C, Grimes J~P and Davidsen A~F 1997 \apj\ {\bf 475} 469 
\end{thereferences}
\end{document}